\documentclass{article}
\usepackage{spconf,amsmath,graphicx}
\usepackage{amsfonts,amssymb}

\usepackage{booktabs}
\usepackage{multirow}
\usepackage{tikz}
\usepackage{url}
\usepackage{subfig}

\title{Don't shoot butterfly with rifles: Multi-channel CONTINUOUS SPEECH SEPARATION WITH EARLY EXIT Transformer}
\name{Sanyuan Chen, Yu Wu,  Zhuo Chen, Takuya Yoshioka, Shujie Liu, Jinyu Li\sthanks{Emails: v-sanych, yuwu1, zhuc, tayoshio, shujliu, jinyli @microsoft.com}}
\address{Microsoft Corporation}
	
\begin{document}
\ninept
\maketitle
\begin{abstract}
With its strong modeling capacity that comes from a multi-head and multi-layer structure, Transformer is a very powerful model for learning a sequential representation and has been successfully applied to speech separation recently. However, multi-channel speech separation sometimes does not necessarily need such a heavy structure for all time frames especially when the cross-talker challenge happens only occasionally. For example, in conversation scenarios, most regions contain only a single active speaker, where the separation task downgrades to a single speaker enhancement problem. 
It turns out that using a very deep network structure for dealing with signals with a low overlap ratio not only negatively affects the inference efficiency but also hurts the separation performance.
To deal with this problem, we propose an early exit mechanism, which enables the Transformer model to handle different cases with adaptive depth. Experimental results indicate that not only does the early exit mechanism accelerate the inference, but it also improves the accuracy. 
\end{abstract}
\section{Introduction}
\label{sec:intro}

Speech separation plays a vital role in  front-end speech processing, aiming to handle the cocktail party problem.  Starting from deep clustering (DC) \cite{hershey2016deep,isik2016single} and permutation invariant training (PIT) \cite{yu2017permutation, kolbaek2017multitalker}, a variety of separation models have been shown effective in separating overlapped speech \cite{yoshioka2019advances,yoshioka2019low}. Recently, the deep learning methods have been rigorously explored for better speech separation capability, including dual-path RNN \cite{luo2020dual}, Conv-tasnet \cite{luo2019conv}, and deep CASA \cite{shi2020furca} employing RNN and CNN structures. With the success of Transformer model in speech community \cite{dong2018speech, Li2020Comparison}, Transformer \cite{chang2020end} and its variants \cite{chen2020continuous2} have successfully been applied to this task.

The Transformer model integrates a stack of self-attention layers to model the speech representation. Prior work shows that a deeper structure yields superior performance \cite{pham2019very}. For example, for automatic speech recognition (ASR) tasks, a common setting is to use twelve \cite{karita2019comparative} or more layers \cite{wang2019semantic} in the encoder. 
However, continuous speech separation (CSS), which we are addressing, is a simpler task especially with a multi-channel setting. Multiple microphones combine to provide rich spatial information that  allows simple models to perform the separation job with high accuracy. Applying a deep Transformer for the multi-channel CSS might be overkill for frames with only one active speaker, resulting in two problems:  
1) Real-time inference is usually preferred for product deployment especially for resource-constrained devices. The Transformer has a \textbf{heavy runtime cost} due to its deep encoder. Hence, it is necessary to speed up the execution of the Transformer-based speech separation models. 
2) The Transformer model may suffer from the ``\textbf{overthinking}'' problem \cite{kaya2019shallow} as it contains too many encoder layers. We assume that a shallow Transformer encoder is sufficient to handle the non-overlapped speech well and that a deep Transformer model could potentially degrade the speech estimation. 

Inspired by the depth-adaptive inference method \cite{kaya2019shallow}, we propose to mitigate these problems with an Early Exit mechanism, which essentially makes predictions at an earlier layer for less overlapped speech while using higher layers for speech with high overlap rate. 
We believe that the first few layers are sufficient to handle the less overlapped speech and thus an early exit scheme reduces the overall runtime cost. 
When the input contains a lot of overlaps, higher layers are automatically triggered to perform more complex analysis and generate more accurate separation results. Specifically, we introduce a mask estimator to each transformer layer and dynamically stop the inference if the predictions from two consecutive layers are sufficiently similar, based on the normalized Euclidean distance of the two prediction matrices. 

We conduct experiments on the LibriCSS dataset \cite{chen2020continuous}. The experimental results show that a stricter threshold (hard to exit) leads to the better performance on large-overlapped utterances and worse performance on the small-overlapped utterances, which is consistent with our intuition. With threshold tuning, the proposed model improves separation quality of the small-overlapped speech while keeping the performance on large-overlapped ones. Moreover, the early exit mechanism enables the Transformer model to achieve better separation performance while 2x the inference speed.

\label{sec:approach}

\begin{figure*}[t]		
	\begin{center}
		\includegraphics[width=1.8\columnwidth]{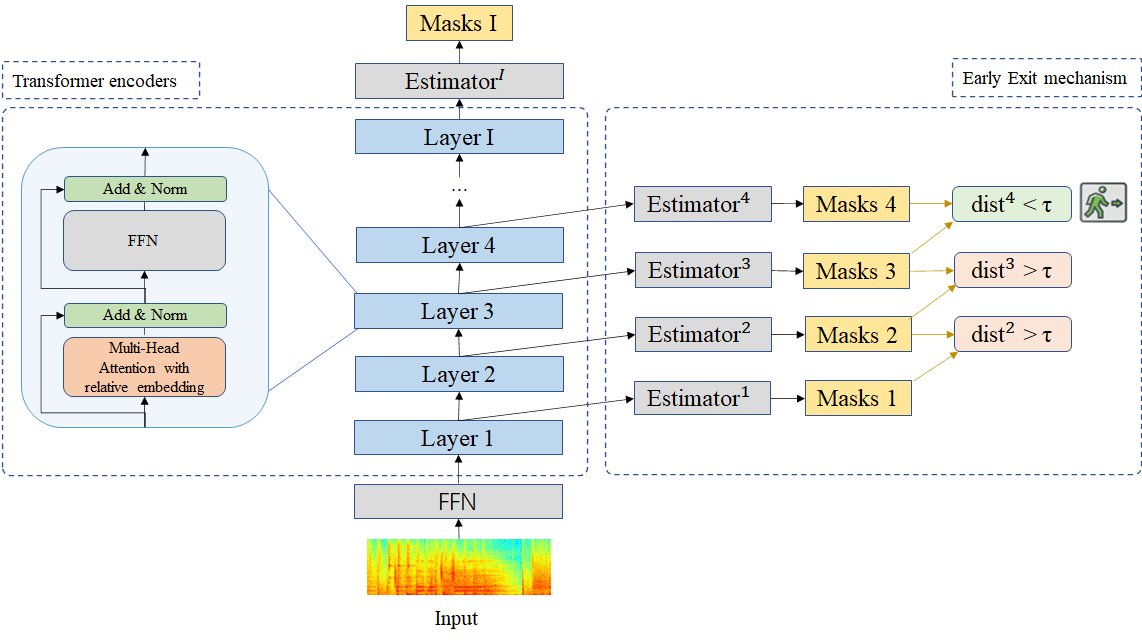}
	\end{center}
	\vspace{-1em}
	
	\caption{The architecture of Early Exit Transformer. We attach a mask estimator to each Transformer encoder layer and dynamically stop inference if the predictions are similar between two consecutive layers.}\label{fig:early_exit_transformer} 
    \label{ssec:model}
\end{figure*} 

\section{Approach}

\subsection{Problem Formulation}
Given a continuously provided signal including multiple talkers, 
CSS aims to retrieve individual constituting utterances and route them to one of its output channels in such a way that each output signal no longer contains overlapped utterances. 
This is typically performed by applying a sliding window to the input signal and performing the separation at each window position. 
Within each window, a separation model generates a fixed number of outputs.

Let $y(t)$ denote the mixed signal and $x_s(t)$ the $s$-th individual target signal, where $t$ is the time index. The mixed signal is modeled as follows: 
\begin{equation}
    y(t)=\sum_{s=1}^S x_s(t). 
\end{equation} 
We also denote 
their short-time Fourier transforms (STFTs) as $\mathbf{Y}(t,f)$ and $\mathbf{X}_s(t,f)$,  respectively. $f$ denotes frequency domain.

When $C$ microphones are available, 
the model input consists of a concatenation of the STFT features of the first channel and the inter-channel phase difference between the $i$-th channel and the first channel. Thus, the features may be represented as 
\begin{equation}
    \mathbf{Y}(t,f) = \mathbf{Y}^1(t,f) \oplus \text{IPD}(2) \ldots \oplus \text{IPD}(C) \
\end{equation} where $\mathbf{Y}^i(t,f)$ denotes the STFT of the $i$-th channel, $ \text{IPD}(i) = \theta^{i}(t,f) - \theta^1(t,f)$, and $\theta^{i}(t,f)$ is the phase of $\mathbf{Y}^i(t,f)$. Each feature dimension is normalized along the time axis. 

Following \cite{wang2014training, erdogan2017deep}, instead of directly computing the STFT of the individual signals $[\mathbf{X}_1(t,f) \ldots \mathbf{X}_S(t,f)]$, we estimate a group of masks $\mathbf{M}(t,f) = [\mathbf{M}_1(t,f) \ldots \mathbf{M}_S(t,f)]$ with a deep learning model $f(\cdot)$.  
Then, for the $s$-th individual signal, $\mathbf{X}_s(t,f)$ is obtained either by beamforming or by masking, i.e., $\mathbf{M}_s(t,f) \odot  \mathbf{Y}^1(t,f)$ where $\odot$ is the elementwise product. 

In the following section, we will first introduce the Transformer model for speech separation and then our Early Exit Transformer network.

\subsection{Transformer Model}

As shown in the left side of Figure~\ref{fig:early_exit_transformer}, We estimate the masks from the input mixed signals with the Transformer model \cite{vaswani2017attention} which is composed of a stack of identical encoder layers. Each layer consists of a multi-head self-attention module and a position-wise fully connected feed-forward module. 

The input of the Transformer model $\mathbf{h}_{0}$ is a linear conversion of the input $\mathbf{Y}(t,f)$ with a feed-forward module  $\text{FFN}(\cdot)$: 
\begin{flalign}
\mathbf{h}_{0} & =\text{FFN}(\mathbf{Y}(t,f)). 
\end{flalign}

Given the input, $\mathbf{h}_{i-1}$, of the $i$-th layer, the output $\mathbf{h}_{i}$ is calculated as
\begin{flalign}
\mathbf{h}_{i}' & =\text{layernorm} (\mathbf{h}_{i-1} + \text{MultiHeadAttention} (\mathbf{h}_{i-1}))  \\
\mathbf{h}_{i} &= \text{layernorm} (\mathbf{h}_{i}' + \text{FFN}(\mathbf{h}_{i}')), 
\end{flalign} where
$\text{MultiHeadAttention}(\cdot)$ and $\text{layernorm}(\cdot)$ denote the multi-head self-attention module and the layer normalization, respectively.

The multi-head self-attention module is implemented with relative position embedding as follows:

\begin{flalign} 
\text{Multihead}(\mathbf{\mathbf{h}_{i-1}}) &= [\mathbf{H}_1 \ldots \mathbf{H}_{d_{head}}]\mathbf{W}^{head}\\
 \text{where~}  \mathbf{H}_j &=\text{softmax}\bigg(\frac{\mathbf{Q}_j(\mathbf{K}_j+\mathbf{pos})^\intercal}{\sqrt{d_k}}\bigg)\mathbf{V}_j, \label{self_att} \\ 
  \nonumber
\end{flalign} where  
$d_k$ is the hidden layer dimensionality, and $d_{head}$ is the number of attention heads. $\mathbf{Q_j},\mathbf{K_j},\mathbf{V_j}$ are linear conversions of the input $\mathbf{h}_{i-1}$ with different parameter matrices.
$\mathbf{pos} = \{rel_{m,n}\} \in \mathbb{R}^{M \times M \times d_k}$ is the relative position embedding \cite{shaw2018self}, where $M$ is the maximum chunk length, and $rel_{m,n} \in \mathbb{R}^{d_k}$ represents the offset of the $m$-th vector in $\mathbf{Q_i}$ and the $n$-th vector in $\mathbf{K_i}$. 

Given the output, $\mathbf{h}_{I}$, of the final layer, we obtain the masks $\mathbf{M}(t,f)$ with $\text{Estimator}^I(\cdot)$, an estimator  consisting of a feed-forward module  and a sigmoid activation function, i.e., 
\begin{flalign} 
    \mathbf{M}(t,f) &= \text{Estimator}^I(\mathbf{h}_{I}) \\
    &= \text{sigmoid}( \text{FFN}(\mathbf{h}_{I})). 
\end{flalign}

\subsection{Early Exit Transformer}
\label{ssec:early_exit}

Despite the promising performance, the Transformer model with deep layers is prone to ``\textbf{heavy runtime cost}'' and ``\textbf{overthinking}'' in the speech separation task.
To overcome this, based on the assumption that the first few layers are sufficient to handle less overlapped speech, we propose an Early Exit Transformer model (see Fig.~\ref{fig:early_exit_transformer}) to estimate the masks by dynamically choosing the number of layers to use. 
Specifically, we attach a layerwise estimator,  $\text{Estimator}^i(\cdot)$, to the output of each Transformer encoder layer $\mathbf{h}_{i}$, based on which, we can predict the masks $\mathbf{M}^{i}(t,f)$ at each internal layer:
\begin{flalign} 
    \mathbf{M}^{i}(t,f) &= \text{Estimator}^i(\mathbf{h}_{i}) \\
    &= \text{sigmoid}( \text{FFN}(\mathbf{h}_{i})).\\\nonumber
\end{flalign}
During the inference, given the output of the $i$-th layer with $i > 1$, we calculate the normalized Euclidean Distance $\text{dist}^i$ between the estimated masks of the $(i\text{-1})$-th layer and the $i$-th layer: 
\begin{equation}
     \text{mean}_{t,f}\bigg( \text{EuclideanDistance}\big(\mathbf{M}^{i-1}(t,f), \mathbf{M}^{i}(t,f)\big) \bigg) \\
      \end{equation}
Given a pre-defined threshold $\tau$, if $\text{dist}^i < \tau$ for the two consecutive layers, we terminate the inference process and output $\mathbf{M}^{i}(t,f))$ as the final prediction masks.
Instead of performing the inference using all the encoder layers, the early exit mechanism makes the predictions with the first few layers for the small-overlap segments, which can accelerate the inference process and potentially reduce the ``overthinking'' problem.

During the training, besides the parameters for the Transformer model, the Estimators attached to the internal layers are also trained to predict the masks from the hidden states. 
Therefore, for each Estimator, we apply PIT \cite{yu2017permutation, kolbaek2017multitalker} to minimize $\text{Loss}^i$ which is the Euclidean distance between the reference and the mask predicted by $\text{Estimator}^i(\cdot)$. The final loss is the weighted average function, following \cite{kaya2019shallow}, as
\begin{equation}
    \text{Loss} = \frac{\sum_{i=1}^I i \cdot \text{Loss}^i}{\sum_{i=1}^I i}
\end{equation} where $\frac{i}{\sum_{i=1}^I i}$ is used as the weight for the loss of the $i$-th estimator, Estimator$^i$. A deeper layer is assigned with a larger weight in the loss computation. The  intuition behind this is that the more complex the model becomes, the more sensitive it gets to prediction errors. Moreover, while the first layer receives the gradients from all layers, only the gradients of $[i,I]$ layers  back-propagate to the $i$-th layer. Thus, giving a larger loss weight to a deeper layer stabilizes the training process.

\begin{table*}[t]
    \centering
    \caption{ Utterance-wise evaluation. Two numbers in a cell denote \%WER of the \textbf{hybrid SR model} used in LibriCSS \cite{chen2020continuous} and \textbf{end-to-end transformer} based SR model \cite{wang2019semantic}. 0S: 0\% overlap
with short inter-utterance silence. 0L: 0\% overlap with a long inter-utterance silence.}
    \label{tab:7ch_utt_result}        
    \vspace{-1em}
    \setlength{\tabcolsep}{1.5mm}
    \begin{tabular}{l|cc|cccccc} 
         \toprule
		\multirow{2}{*}{\textbf{System}} &
		\textbf{Avg. exit} & \textbf{Speed-} &
		\multicolumn{6}{c}{\textbf{Overlap ratio in \%}} \\ 
		& \textbf{layer} & \textbf{up} &
		0S & 0L & 10 & 20 & 30 & 40   \\ 
		\midrule
		No separation \cite{chen2020continuous} & - & - & 11.8/5.5 & 11.7/5.2 & 18.8/11.4 & 27.2/18.8 & 35.6/27.7 & 43.3/36.6 \\
		BLSTM \cite{chen2020continuous2} & - & -  & \textbf{7.0}/\textbf{3.1} & 7.5/\textbf{3.3} & 10.8/4.3 & 13.4/5.6 & 16.5/7.5 & 18.8/8.9 \\ 
		Transformer \cite{chen2020continuous2} & 16.0 & $1.00\times$ & 8.3/3.4 & 8.4/3.4 & 11.4/4.1 & 12.5/\textbf{4.8} & 14.7/6.4 & 16.9/7.2 \\	
		\midrule
		Early Exit Transformer ($\tau = 0$)  & 16.0 & $0.92\times$ & 8.9/3.4 & 9.4/3.6 & 12.3/4.2 & 14.7/5.0 & 15.1/\textbf{6.2} & \textbf{16.5}/\textbf{6.6} \\
        Early Exit Transformer ($\tau = 8e-5$)  & 6.9 & $2.60\times$ & 7.6/\textbf{3.2} & 7.7/\textbf{3.3} & 10.1/\textbf{3.8} & 12.4/\textbf{4.8} & \textbf{14.4}/\textbf{6.2} & \textbf{16.4}/6.9 \\
        Early Exit Transformer ($\tau = 1.5e-4$)  & 4.8 & $4.08\times$ & 7.8/\textbf{3.2} & 7.6/\textbf{3.4} & \textbf{9.8}/\textbf{3.8} & \textbf{12.2}/5.1 & 14.7/6.7 & 17.9/7.8 \\
        Early Exit Transformer ($\tau = \infty$)  & 2.0 & $6.59\times$ & \textbf{7.1}/\textbf{3.1} & \textbf{7.3}/\textbf{3.3} & 10.0/4.4 & 13.6/6.1 & 17.0/8.4 & 20.5/10.4 \\
		\bottomrule
    \end{tabular}
\end{table*}

\begin{table*}[t]
    \centering
    \caption{Continuous speech separation evaluation}
    \vspace{-1em}
    \label{tab:7ch_cont_result}    \setlength{\tabcolsep}{1.5mm}
    \begin{tabular}{l|cc|cccccc} 
         \toprule
		\multirow{2}{*}{\textbf{System}} &
		\textbf{Avg. exit} & \textbf{Speed-} &
		\multicolumn{6}{c}{\textbf{Overlap ratio in \%}} \\ 
		& \textbf{layer} & \textbf{up} &
		0S & 0L & 10 & 20 & 30 & 40   \\ 
		\midrule
		No separation \cite{chen2020continuous} & - & - & 15.4/12.7 & 11.5/5.7 & 21.7/17.6 & 27.0/24.4 & 34.3/30.9 & 40.5/37.5 \\
		BLSTM  \cite{chen2020continuous2} & - & - & 11.4/6.0 & \textbf{8.4}/\textbf{4.1} & 13.1/7.0 & 14.9/7.9 & 18.7/11.5 & 20.5/12.3 \\ 
		Transformer  \cite{chen2020continuous2} & 16.0 & $1.00\times$ & 12.0/5.6 & 9.1/4.4 & 13.4/6.2 & 14.4/\textbf{6.8} & 18.5/9.7 & 19.9/\textbf{10.3} \\
		\midrule
		Early Exit Transformer ($\tau = 0$)  & 16.0 & $0.76\times$ & 14.1/6.2 & 10.3/4.6 & 17.2/7.1 & 17.3/7.5 & 23.0/10.8 & 23.5/12.0 \\
		Early Exit Transformer ($\tau = 1e-4$)  & 7.5 & $1.47\times$ & \textbf{11.3}/5.4 & 8.9/4.4 & \textbf{12.7}/\textbf{6.0} & \textbf{13.8}/\textbf{6.7} & 17.8/\textbf{9.3} & \textbf{19.7}/10.5 \\
        Early Exit Transformer ($\tau = 1.5e-4$)  & 5.8 & $1.88\times$ & 11.5/\textbf{5.2} & 8.9/4.3 & \textbf{12.6}/\textbf{6.0} & \textbf{13.7}/6.9 & \textbf{17.6}/9.5 & \textbf{19.6}/\textbf{10.3} \\
        Early Exit Transformer ($\tau = 2e-4$)  & 5.2 & $2.08\times$ & \textbf{11.2}/5.6 & 8.8/4.5 & \textbf{12.7}/6.3 & 13.9/7.2 & 18.5/9.5 & \textbf{19.6}/10.9 \\
        Early Exit Transformer ($\tau = \infty$)  & 2.0 & $4.74\times$ & 14.7/14.6 & 8.7/6.9 & 16.1/13.7 & 17.8/15.2 & 22.5/18.2 & 24.8/18.9 \\
		\bottomrule
    \end{tabular}
\end{table*}

\section{experiment}
\label{sec:experiment}

\subsection{Datasets}
We train the models with 219 hours of artificially reverberated and mixed speech signals  sampled randomly from WSJ1 \cite{wsj1}. Following \cite{yoshioka2018multi}, we include four different mixture types in the training data. Each training mixture is generated by randomly picking one or two speakers from the WSJ1 dataset and convolving each with a 7 channel room impulse response (RIR) simulated with the image method \cite{habets2006room}. Then, we rescale and combine them with a source energy ratio between -5 and 5 dB. Simulated isotropic noise \cite{habets2007generating} is also added  at a 0--10 dB signal to noise ratio. The average overlap ratio of the training set is around 50\%.

We evaluate the models on the LibriCSS dataset \cite{chen2020continuous}, which consists of 
10 hours of concatenated and mixed LibriSpeech utterances played and recorded in a meeting room. 
We test our model performance under a seven-channel setting. We conducted both the utterance-wise evaluation and continuous input evaluation (refer to \cite{chen2020continuous} for the two evaluation schemes).

\subsection{Implementation Details}
Our baseline speech separation models are BLSTM and vanilla Transformer.
The BLSTM model consists of three BLSTM layers with 1024 input dimensions and 512 hidden dimensions, resulting in  21.80M parameters.
Three sigmoid projection layers are appended to estimate three masks, two for speakers and one for noise. 
We use the Adam optimizer \cite{kingma2014adam} to train the BLSTM model with the learning rate initialized to 1e-3. The learning rate is decreased by half if the validation loss stops decreasing for 2 epochs. Training is performed for 100 epochs.
The Transformer model consists of 16 Transformer encoder layers with 4 attention heads, 256 attention dimensions and 2048 FFN dimensions, resulting in 21.90M parameters.
We use the AdamW optimizer \cite{loshchilov2018decoupled} to train the Transformer model with the weight decay set to 1e-2.
The learning rate is 1e-4 and the warm-up learning schedule with linear decay is used, where the warm-up step is 10,000, and the training step is 260,000. 

Our Early Exit Transformer model is implemented with the same Transformer encoders as the baseline Transformer model.
The model is optimized with the weighted average loss (as described in Section~\ref{ssec:early_exit}) with the same hyperparameters as the baseline.
During inference, we vary the early exit threshold in \{0, 3e-5, 5e-5, 8e-5, 1e-4, 1.5e-4, 2e-4, $\infty$\} to control the exit layer and thus the speed-up ratio.
We evaluate the speech separation accuracy with two ASR models. One is a hybrid system with a BLSTM based acoustic model and a 4-gram language model as used in the original LibriCSS paper \cite{chen2020continuous}. The other is one of the best open source end-to-end transformer \cite{wang2019semantic} based ASR models \footnote{https://github.com/MarkWuNLP/SemanticMask} which achieves WERs of 2.08\% and 4.95\%
for LibriSpeech test-clean and test-other, respectively. 
As with \cite{chen2020continuous}, by leverating the multiple microphones, the individual target signals are generated with 
mask-based adaptive minimum variance distortionless response (MVDR) beamforming.

\subsection{Evaluation Results}
\label{ssec:utt_eval}

Table~\ref{tab:7ch_utt_result} shows the WERs of our Early Exit Transformer with different threshold $\tau$ values as well as those of  the baselines for the utterance-wise evaluation.
With a larger threshold, the inference process tended to exit at a lower layer, and greater speed-up was obtained. 
We also found the performance on the low overlap ratio sets benefited from the use of fewer inference layers, which implies the ``overthinking'' problem of the vanilla Transformer model.
Specifically, when $\tau = \infty$, the inference process always halted at the second layer. This yielded a $6.59\times$ speed-up and achieved the best WERs for the two non-overlap settings.
The use of a smaller threshold led to a better separation performance for high overlap ratio settings. When $\tau = 0$, its performance degraded on the small overlap ratio sets. This may have been caused by the mismatch between the training and inference, i.e.,  with the proposed method, the model tries to predict the mask correctly at all the layers during while only the last layer's result is used at the inference time. Moreover, it is slower than the vanilla Transformer since every layer predicts the output once. 
With tuned threshold, better results were obtained by mitigating the ``overthinking'' problem.
With $\tau=1.5e-4$, our Early Exit Transformer  achieved better results in the small overlap ratio settings than the vanilla Transformer while achieving a $4.08\times$ speed-up.
With a modest threshold setting ($\tau=8e-5$), the inference time was halved while also achieving better speech separation performance with both of the two ASR models for all overlap settings.

Table~\ref{tab:7ch_cont_result} shows 
the continuous evaluation results. 
As with the utterance-wise evaluation, the Early Exit Transformer models achieved superior performance to the vanilla Transformer while improving the inference time by a factor of two. The improvements were more prominent on the small-overlapped test sets. 
In contrast to the utterance-wise speech separation, we observed that the inference process tended to stop at a higher layer for the same threshold and that the best results for the non-overlap settings were achieved with some internal layer rather than with the second layer.
This could be because, in CSS, each model evaluation used a shorter chunk than the typical utterance length of the utterance-wise evaluation dataset, making the task harder.

\subsection{Discussion and Analysis}
In addition to the main experiments, there are several interesting questions that should be discussed. 

\begin{figure}[h]
	\centering
	{\label{sfig:a}\includegraphics[scale=0.45, trim={0.5cm 0cm 1.5cm 1.3cm}, clip]{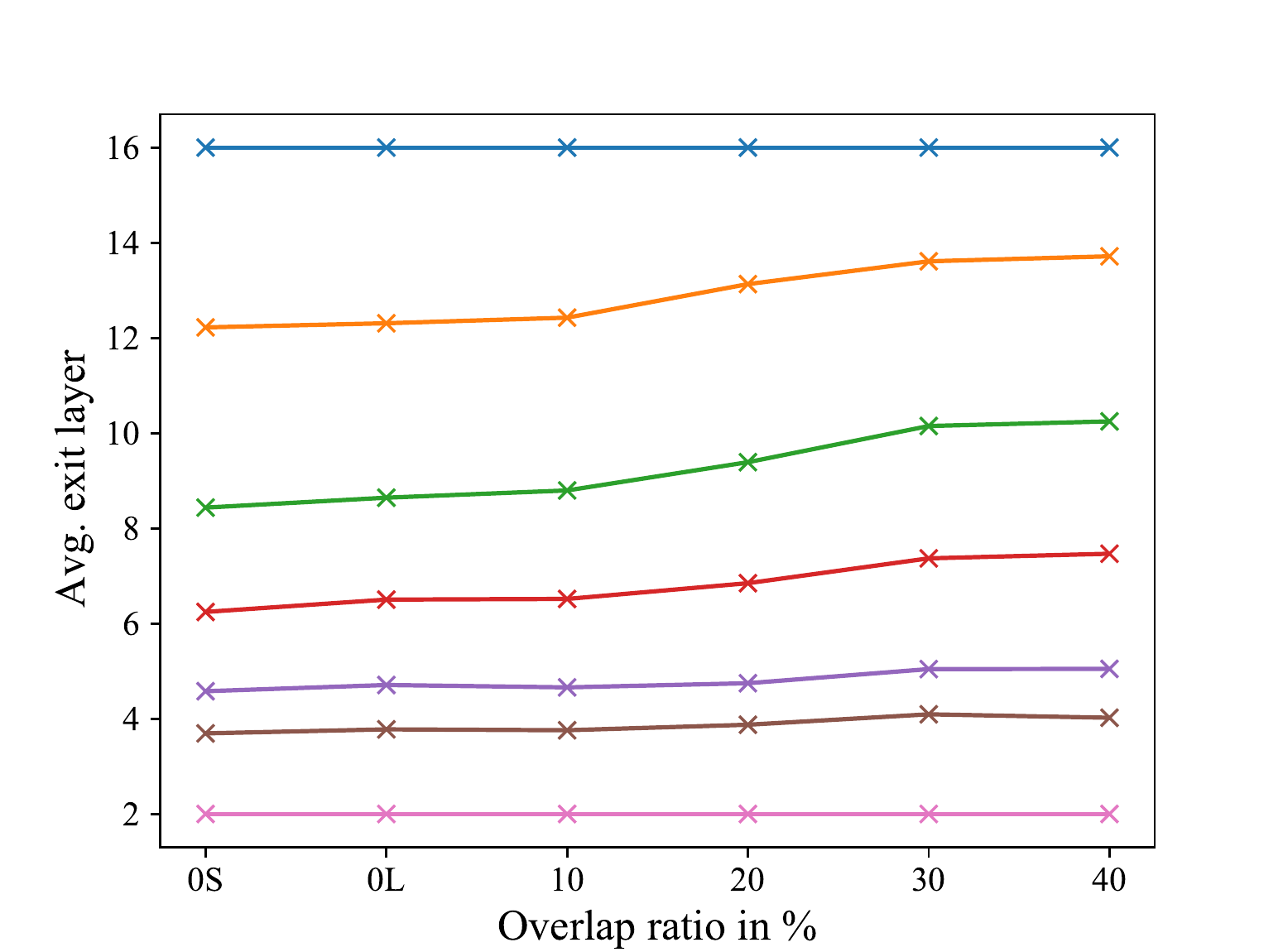}}\quad
	{\label{sfig:l}\includegraphics[scale=0.55, trim={0cm 0.5cm 0cm 1cm}, clip]{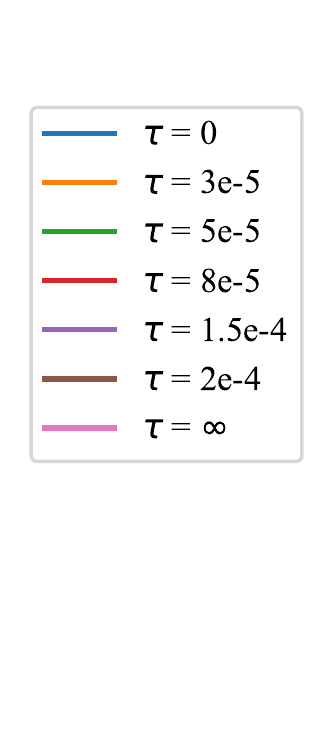}}\quad
	\caption{The average exit layer of Early Exit Transformer across different testsets with different threshold $\tau$ for the utterance-wise evaluation.
	}\label{fig:layer_wer_analysis}
\end{figure}

\textbf{Exit layer across different testsets: } We first explore what the exit layer distribution is with respect to different overlapped ratios. Figure \ref{fig:layer_wer_analysis} shows that the averaged exit layer slightly increases as the overlap ratio becomes larger. 
When $\tau=8e-5$, Early Exit Transformer makes predictions  one layer deeper on 40$\%$ overlap testset compared to 0S testset. The increment is more significant with $\tau=5e-5$,
demonstrating that small overlapped cases tend to exit at shallow layers which is consistent with our intuition.  

\textbf{Performance for the single channel scenario: } We also tried the early exit mechanism on single channel speech separation task, but obtained negative results. For single channel scenario, the conclusion is the more layers we use, the better performance we get. We think that speech separation for single channel is much more challenging due to the absence of the microphone array signal, and less than 16 layers are not enough to handle this task well. We leave it for future work to explore the early exit mechanism for the single channel. 

\label{ssec:discussion}

\section{conclusion}
\label{sec:conclusion}
We elaborate an early exit mechanism for Transformer based multi-channel speech separation, which aims to address the ``over-thinking"  problem and accelerate inference stage simultaneously. Each Transformer layer is equipped with a mask estimator, and the early exit is triggered if the outputs of two successive layers are similar. Experiment results show that it does not only speed up inference, but also improves the performance on small-overlapped testsets, which is consistent with our intuition. Regarding single-channel evaluation, we observe negative result since the task is too challenging to handle. In the future, we will study speed up Transformer based separation model from other perspectives. 

\vfill\pagebreak

\bibliographystyle{IEEEbib}
\bibliography{strings,refs}

\end{document}